\documentclass[11pt,a4paper,final]{article}

\usepackage[a4paper,left=3cm,right=3cm,top=3cm,bottom=3cm]{geometry}

\usepackage[utf8]{inputenc}
\usepackage[USenglish]{babel}

\usepackage{amsmath}
\usepackage{amsfonts}
\usepackage{amssymb}
\usepackage{graphicx}
\usepackage{xcolor}
\usepackage{setspace}
\usepackage{hyperref}
\usepackage{siunitx}

\newcommand{\dext}{\text{d}}
\newcommand{\GN}{G_\text{N}}

\usepackage{hyperref}
\hypersetup{
	pdftitle={Shear transport far from equilibrium via holography},
	pdfauthor={Michael Florian Wondrak, Matthias Kaminski, Marcus Bleicher},
	pdffitwindow=true,
	colorlinks=true,
	linkcolor={red!50!black},
	citecolor={blue!50!black},
	urlcolor={blue!80!black}
}

\usepackage[comma,numbers,sort&compress]{natbib}
\begin{document}

\begin{center}
\begin{spacing}{1.5}
{\Large\bf Shear transport far from equilibrium via holography}
\end{spacing}
\end{center}

\vspace{-0.1cm}
\begin{center}
\textbf{Michael~F.~Wondrak}$^{a,}$\footnote{\texttt{wondrak@itp.uni-frankfurt.de}}
\textbf{Matthias~Kaminski}$^{b,}$\footnote{\texttt{mski@ua.edu}}
\textbf{and Marcus~Bleicher}$^{a,}$\footnote{\texttt{bleicher@th.physik.uni-frankfurt.de}}

\vspace{.6truecm}
{\em $^b$Institut f\"ur Theoretische Physik,\\
Johann Wolfgang Goethe-Universit\"at Frankfurt am Main,\\
Max-von-Laue-Stra\ss{}e 1, 60438 Frankfurt am Main, Germany}\\

\vspace{.3truecm}
{\em $^b$Department of Physics and Astronomy, University of Alabama, \\
514 University Blvd., Tuscaloosa, AL 35487, USA}\\

\vspace{.6truecm}
February 26, 2020
\end{center}

\vspace{0.1cm}

\begin{abstract}
\noindent{\small%
\noindent In heavy-ion collisions, the quark-gluon plasma is produced far from equilibrium.
This regime is currently inaccessible by quantum chromodynamics~(QCD) computations.
We calculate shear transport and entropy far from equilibrium in a holographic model, defining a time-dependent ratio of shear viscosity to entropy density, $\eta/s$.
Large deviations of up to 60\% from its near-equilibrium value, $1/4\pi$, are found for realistic situations at the Large Hadron Collider.
We predict the far-from-equilibrium time-dependence of $\eta/s$ to substantially affect the evolution of the QCD plasma and to impact the extraction of QCD properties from flow coefficients in heavy-ion collision data.

}
\end{abstract}

\clearpage


\section{\label{sec:intro}Introduction}

Non-equilibrium systems are abundant in Nature and remain elusive despite various theoretical approaches~\cite{Lieb:1999,Lebon2008,Khemani:2019nzi}. Quantum systems far from equilibrium pose a harder problem yet. 
Such systems are, for instance, the rapidly expanding early universe, the quark-gluon plasma~(QGP) generated in heavy-ion collisions~\cite{Romatschke:2017ejr}, 
or condensed matter experiments in which external parameters are rapidly changed (quench)~\cite{Calabrese:2006rx}.

A hallmark of non-equilibrium systems is entropy production. Entropy production is a fundamental measure for time evolution. It occurs in conjunction with dissipative processes, for example the transfer of momentum in the direction transverse to the fluid flow, called shear transport. In hydrodynamics, the ability of a fluid for shear transport is quantified by the shear viscosity, $\eta$. In plasmas, the ratio of shear viscosity to entropy density, $\eta/s$, is a key property indicating how far a plasma deviates from an ideal fluid.

In this work, the central question is to what extent strong time-dependence
and far-from-equilibrium physics 
affect shear transport in the QGP. That is, how large are the corrections to $\eta/s$?

Perturbative methods have successfully provided values for the QGP shear viscosity near equilibrium~\cite{Arnold:2000dr,Arnold:2003zc}. However, the  corrections at leading and next-to-leading order in the quantum chromodynamics~(QCD) coupling constant are large~\cite{Ghiglieri:2018dib}.
Hence, the challenge is much harder as the QGP is strongly coupled during much of its time evolution~\cite{Gyulassy:2004zy}, a regime inaccessible to all perturbative methods. 

Access to this question was granted with the advent of the gauge/gravity duality (aka holography, or AdS/CFT)~\cite{Maldacena:1997re}. For a generic plasma at strong coupling, an astonishingly low value of $\eta/s=1/4\pi \times \hbar/k_\text{B}$~\cite{Policastro:2001yc,Kovtun:2004de} was predicted~\footnote{In the following we apply the so-called natural units by setting $\text{c} \equiv \hbar \equiv k_\text{B} \equiv 1$. Note that we explicitly keep the Newton gravitational constant $\GN$.},~%
\footnote{
The ratio $\eta/s$ has a lower bound in all known systems when taking the hydrodynamic limit~\cite{Kovtun:2004de}. 
The value of that bound depends on the system~\cite{Kats:2007mq,Brigante:2008gz}. 
In our model, the near-equilibrium value is $\eta/s=1/4\pi$. 
}.
Remarkably, this value was later experimentally found to be consistent with heavy-ion collision data~\cite{Gyulassy:2004zy} in conjunction with hydrodynamic modeling.
Since then, the ratio $\eta/s$ has profoundly impacted the interpretation of heavy-ion collision data over the past decade~\cite{Romatschke:2017ejr}. Hydrodynamic simulations were improved by the introduction of a small but non-zero value of $\eta/s \leq 2.5/4\pi$~\cite{Song:2010mg,Heinz:2011kt}. 
In particular, the elliptic flow of charged particles was used to estimate the value of $\eta/s$ from the experimental data~\cite{Gyulassy:2004zy,Romatschke:2007mq,Luzum:2008cw,Aad:2010bu,Schenke:2011tv,Luzum:2012wu,Acharya:2020taj,Bernhard:2019bmu}. 

During the early stage, the QCD matter created in a heavy-ion collision is expected to be out of equilibrium, even locally~\cite{Romatschke:2017vte}. This calls for more accurate values for the time evolution of $\eta/s$ as input for hydrodynamic models. Thus far, a temperature-dependent $\eta/s$ has been derived only in equilibrium approaches, most notably from the functional renormalization group (FRG) and lattice QCD~\cite{Christiansen:2014ypa,Astrakhantsev:2017nrs}, or~\cite{Soloveva:2019xph}. 
These results were then applied in hydrodynamic models, yielding predictions for various flow coefficients, $v_n$~\cite{Niemi:2011ix,Gale:2012rq,Karpenko:2015xea,Dubla:2018czx}. 

\begin{figure}[ht!]
\center
\includegraphics[width=0.6\textwidth]{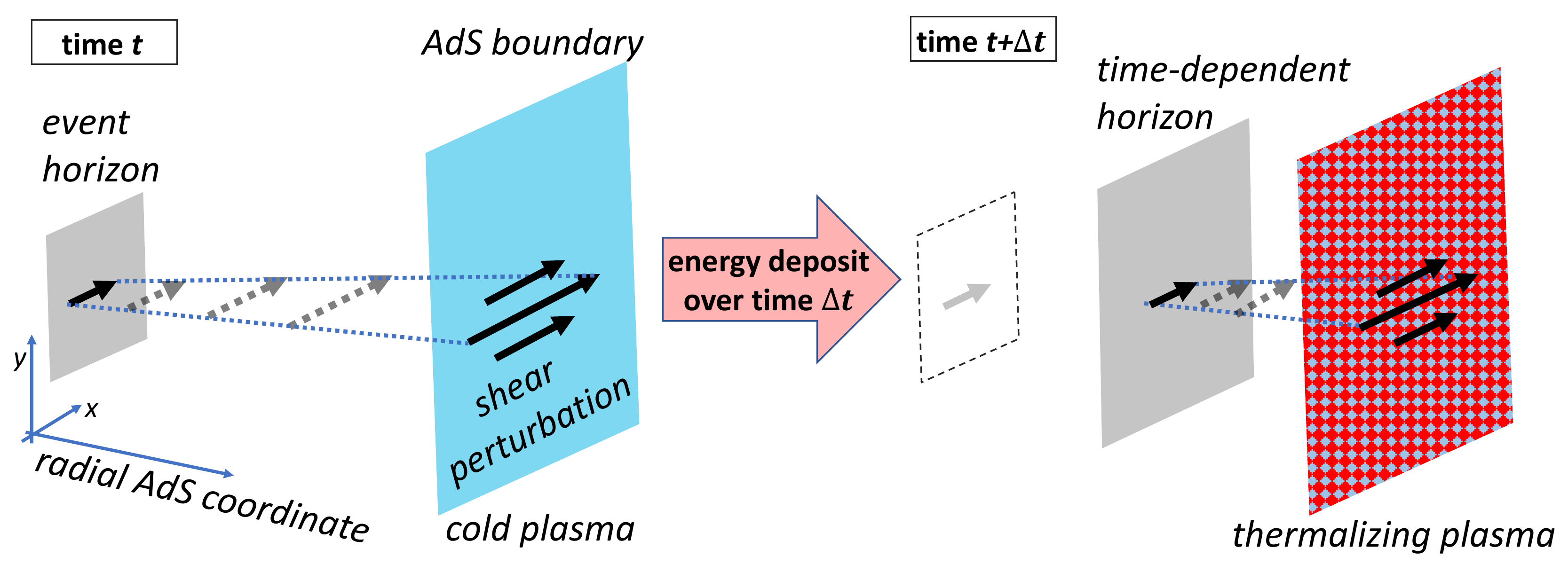}
\caption{\label{fig:introFig}
Holographic setup: 
A rapid energy deposit over a time $\Delta t$ turns a cold plasma into a thermalizing plasma far from equilibrium. 
On the left side, at time $t$, the gravity dual to the cold plasma is a nearly static black brane with a nearly constant event horizon, temperature, and entropy density. 
On the right side, at time $t+\Delta t$, the gravity dual to the thermalizing plasma far from equilibrium is a time-dependent black brane spacetime, from which we  
compute the time-dependent temperature and entropy density of the thermalizing plasma. 
We introduce a shear perturbation and compute the far-from-equilibrium analog of the shear viscosity over entropy density ratio, $\eta/s$. 
}
\end{figure}

In this work, we study the time-dependence of $\eta/s$ in a holographic model far from equilibrium to provide the missing link between the early and the late  collision stage. 
The model is based on a time-dependent charged black brane spacetime in  Einstein-Maxwell theory coupled to time-dependent external sources illustrated in
Fig.~\ref{fig:introFig}~\cite{Vaidya:1951zz,AbajoArrastia:2010yt,Caceres:2012em,Wondrak:2017kgp}.
The infall of matter causes the mass of the black brane to increase rapidly, corresponding to a change of the temperature and the entropy density in the dual field theory. 
To link this to hydrodynamically accessible quantities, we  provide adequate out-of-equilibrium definitions of shear transport, temperature, and entropy.  

We show, for the first time, that large corrections to $\eta/s=1/4\pi$ arise far from equilibrium. 
Varying our model parameters over a large range of values relevant to heavy-ion collisions, we demonstrate that these corrections to $\eta/s$ are of order one in all cases.

\section{\label{sec:holographicModel}Holographic Model}

We employ superconformal Yang-Mills (SYM) theory with $N_\text{c}\to \infty$ degrees of freedom in the limit of large 't Hooft coupling $\lambda$~\cite{Aharony:2008ug} as a well established model for the QCD quark-gluon plasma in the strong coupling regime.
The AdS/CFT correspondence relates it to Einstein-Maxwell theory in an asymptotically Anti-de Sitter (AdS) spacetime with the metric field $g_{mn}$ and the $U(1)$ gauge field $A_m$,
\begin{align}
\label{eq:action}
S 
&= S_\text{EM} + S_\text{matter}\\
S_\text{EM}
&= \int \dext^{4}x\; \sqrt{-g} \left(\frac{1}{16\pi G_\text{N}} \left(R +6\right) 
  -\frac{1}{4}\, F^{mn} F_{mn}\right)\, ,\nonumber
\end{align}
where $R$ is the Ricci scalar associated with the metric $g_{mn}$, which has the determinant $g$, and $F=\dext A$ is the field strength. 
Here, all dimensionful quantities are scaled to the AdS radius.
The model, Eq.~\eqref{eq:action}, includes matter, $S_\text{matter}$, which is to be accreted by the black brane, introducing time-dependence. 

A thermalizing field theory plasma is dual to a time-dependent metric background, i.e.~a dynamic spacetime on the gravity side, see Fig.~\ref{fig:introFig}. 
Such a spacetime can be achieved by 
the infall of matter from the boundary toward the singularity. 
This matter gives rise to the gravitational stress-energy tensor as well as to the $U(1)$ current. 
These act as time-dependent sources in the Einstein-Maxwell equations.
As the mass of the black brane increases, the apparent horizon radius and event horizon radius grow with time, see Fig.~\ref{fig:introFig}. 
This is realized in the charged time-dependent Vaidya black brane~\cite{Vaidya:1951zz,AbajoArrastia:2010yt,Caceres:2012em,Wondrak:2017kgp} with the solution 
\begin{align}\label{eq:metric}
ds^2 
&= \frac{1}{z^2} 
 \left( -f(v,z) \dext v^2 -2\dext v\, \dext z  +\dext x^2 +\dext y^2 \right)\, ,\\
A(v)
&= \left( \mu(v) -Q(v)\, z \right) \dext v\, .
\end{align}
Here, the Eddington-Finkelstein coordinate $v$ denotes the null time. At the AdS boundary, it equals the time in the field theory, $v=t$.
We work with the inverse radial AdS coordinate, $z$, for which the AdS boundary is located at $z=0$, and the singularity at $z\to \infty$. 
The time-dependent horizon position follows from the blackening factor, 
$f(v,z) = 1 -2\GN M(v)\, z^3 +4\pi \GN\, Q(v)^2\, z^4$. 
We fix the location of the apparent horizon to be at $z=1$. This fixes the event horizon location to $z=1$.

The mass and charge densities of the brane, $M$ and $Q$, are $v$-dependent. They can be chosen freely to match a given evolution of the temperature, $T$, and chemical potential, $\mu$. 

For our calculations, we increase the mass by $\Delta M \propto 1 +\tanh(v/\Delta t)$ in a characteristic time scale $\Delta t$.
The largest amount of mass (or charge) is falling into the black brane at a time $v=0$. 
The background solution, Eq.~\eqref{eq:metric}, and thus the energy increase in the field theory, is homogeneous and isotropic along the boundary coordinates.

With the holography-based approach we investigate the dynamics of heavy-ion collisions at the Relativistic Heavy Ion Collider (RHIC) and the Large Hadron Collider (LHC). In this paper, we particularly focus on the earliest stage, the heat-up phase, where the system is the farthest from equilibrium. 
One expects to reach a maximum temperature in the range of $\SI{300}{MeV}$--$\SI{600}{MeV}$ within a time $\SI{.15}{fm}$--$\SI{.6}{fm}$ for central collisions at RHIC, while temperatures of $\SI{600}{MeV}$--$\SI{900}{MeV}$ within similar times at the LHC~\cite{Adare:2009qk}.

\section{Entropy \& Shear Far From Equilibrium} \label{sec:etasFFEq}

In equilibrium, the entropy density of the black brane horizon corresponds to the entropy density of the dual field theory. Fig.~\ref{fig:introFig} illustrates a black brane and its dual field theory on the AdS boundary during an ongoing energy deposit according to the increasing mass function $M(v)$ in $f(v,z)$, Eq.~\eqref{eq:metric}. The black brane changes its surface area, i.e. the black brane entropy changes. 

Out of equilibrium, however, this entropy is not equal to the field theory entropy. Instead, a well-suited measure for the entropy density of the boundary field theory is defined on the boundary itself and can be derived from the on-shell action. The latter follows from Eq.~\eqref{eq:action} by holographic renormalization~\cite{deHaro:2000vlm}. We obtain the generating functional, the pressure $P=M/8\pi$, as a function of the temperature $T$ and the chemical potential $\mu$. 
We define the time-dependent entropy density of the field theory as
$s = {\left(\partial P / \partial T\right)}_\mu$.
The generating functional derived from holography is well-defined in a time-dependent scenario, see {e.g.}~\cite{Skenderis:2008dh}. 
It defines $T(t)$ since holography assures direct access to $P(t)$ and $\mu(t)$. 

Shear in a fluid near- or far-from-equilibrium is encoded in the spatial off-diagonal components of the stress-energy tensor, e.g.~$\langle T^{xy}\rangle$. Near equilibrium, the 2-point function ~$\langle T^{xy} T^{xy}\rangle$ encodes shear transport and is related to the shear viscosity, $\eta$, via a Kubo formula,
\begin{equation}\label{eq:eta}
\eta 
= \lim\limits_{\omega\to 0}\lim\limits_{\bf{k}\to 0} \frac{\langle T^{xy} T^{xy} \rangle (\omega,\bf{k})}{-\text{i}\omega}\, , 
\end{equation}
with Fourier space frequency $\omega$ and momentum $\bf{k}$. 
Far from equilibrium, the shear correlator is well defined in position space. However, time translation invariance is violated and hence plane waves, $\text{e}^{-\text{i}k\cdot x}$, are no solution basis. 
Thus, we work in position space, computing the 2-point function using linear response theory. We introduce a perturbation of the metric, $h^{(0)}_{xy}$, which sources the operator $T^{xy}$. For a localized source, $h^{(0)}_{xy}(\tau) \propto\delta(\tau-t)$, the system response yields the correlator directly,
\begin{eqnarray}\label{eq:TxyTxy}
\left\langle T^{xy}\!\left(t_2
\right) \right\rangle_{\delta(t_1)}
&\propto& -\int\!\dext \tau\; \langle T^{xy} T^{xy} \rangle(\tau,\,t_2) \; 
{\delta (\tau-t_1)} \nonumber\\
&=& - \langle T^{xy} T^{xy} \rangle(t_1,\,t_2)\, ,
\end{eqnarray}
where we suppressed the spatial dependence. Invoking the holographic correspondence, we perturb the metric, Eq.~\eqref{eq:metric}, with a fluctuation $h_{xy}$~\cite{Son:2002sd,Herzog:2002pc,Skenderis:2008dh,Chesler:2013lia}. 
This yields a differential boundary value problem for $h_{xy}$ in $v$ and $z$. It can be cast into an ordinary differential equation in $z$ at an initial time. Two boundary conditions are imposed: a $\delta$-source at $z=0$, and ingoing solutions at $z=1$. The resulting solution is propagated forward in time. 
The solution for $h_{xy}$ at time $t_2$ yields the desired correlator after holographic renormalization~\cite{Banerjee:2016ray,Ishii:2016wwa}.

This correlator can be Wigner transformed to the frequency, $\omega$, which corresponds to the relative time between source and response, $t_2-t_1$. This yields a measure for shear transport far from equilibrium at any average time $t_\text{avg}=(t_1+t_2)/2$,
\begin{equation}\label{eq:etaFFE}
\eta(t_\text{avg})
= \lim\limits_{\omega\to 0} \lim\limits_{\bf{k}\to 0}
\frac{\langle T^{xy} T^{xy}\rangle(\omega,{\bf{k}},t_\text{avg})}{-\text{i}\omega} \, ,
\end{equation}
cf.~\cite{Romatschke:2017vte}. When evaluated in an equilibrium state, Eq.~\eqref{eq:etaFFE} reduces to the known expression for $\eta$ given in Eq.~\eqref{eq:eta}. Since the Wigner transform samples all times, the shear transport measure, Eq.~\eqref{eq:etaFFE}, takes into account the time needed for information to propagate from the horizon to the boundary.

\section{\label{sec:results}Results}
To apply this model to the QGP phase of heavy-ion collisions at RHIC and LHC, we start at the critical temperature of the QGP phase transition, $T_\text{C}=\SI{155}{MeV}$, and raise the temperature by a factor of 2, 4, 6.5, and 10. The duration required for the temperature to increase between $5\%$ and $95\%$ of the peak temperature is referred to as heat-up time. This corresponds to $3.5\, \Delta t$, which we choose to be $\SI{.24}{fm}$ and $\SI{.59}{fm}$, respectively. For now, we consider collisions at vanishing baryochemical potential, $\mu=0$, adequate for collisions at RHIC and~LHC.

The computation of the time-dependent shear viscosity and entropy density allows us to study the ratio $\eta(t_\text{avg})/s(t_\text{avg})$ in the far-from-equilibrium regime. Using the definition of the entropy density provided above, $s$ shows a monotonic increase with time. However, the shear viscosity as a function of time, Eq.~\eqref{eq:etaFFE}, shows an initial dip, then increases monotonically before it approaches the equilibrium value at late times. For some parameter combinations the shear viscosity overshoots the equilibrium value and then asymptotes from above.

%
\begin{figure}[tb]
\center
\includegraphics[width=.5\linewidth]{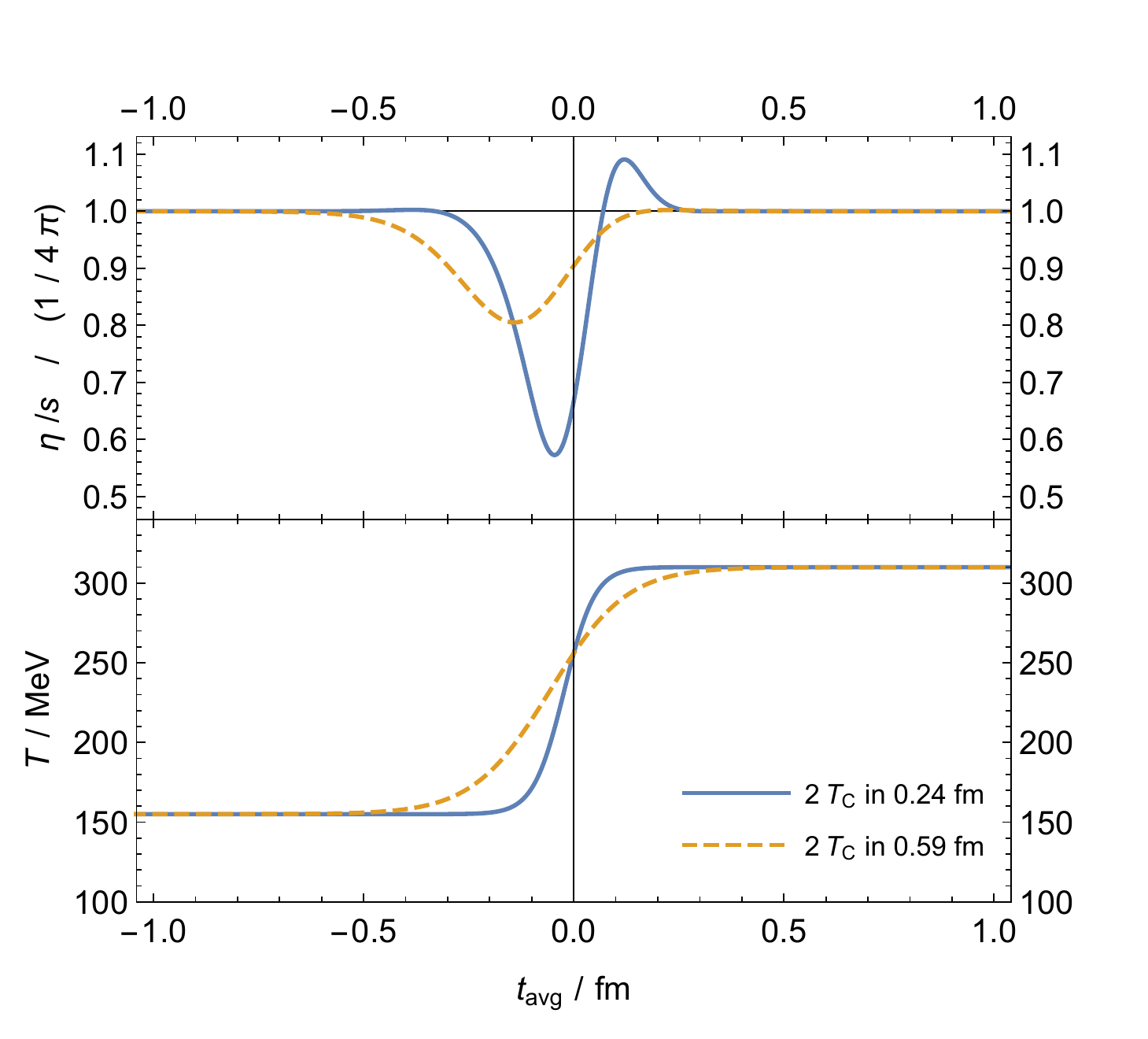}
\caption{\label{fig:etaOfTime}
Time evolution of~$\eta/s$. Lower panel: Two sample temperature profiles $T(t_\text{avg})$ with heat-up to $2T_\text{C}$ within $\SI{.24}{fm}$ and $\SI{.59}{fm}$. Upper panel: $\eta/s$ reacts with a marked minimum. The thin black line represents the near-equilibrium value, $1/4\pi$.
}
\end{figure}
%

Fig.~\ref{fig:etaOfTime} presents explicit examples of the evolution of $\eta/s$ 
together with the associated temperature profiles with peak temperature $2 T_\text{C}$ and two different heat-up times. 
For early and late times, the ratio reaches $1/4\pi$ in agreement with the near-equilibrium value of holographic plasmas dual to Einstein gravity. During the heat-up phase, however, we find a significant decrease with a prominent minimum, lying 20\%--60\% below the equilibrium value.

Both, time and temperature, determine how far the system is driven out of equilibrium: 
Decreasing the heat-up duration or increasing the peak temperature leads to a lower and more pronounced minimum. 

\begin{figure}[tb]
\center
\includegraphics[width=.5\linewidth]{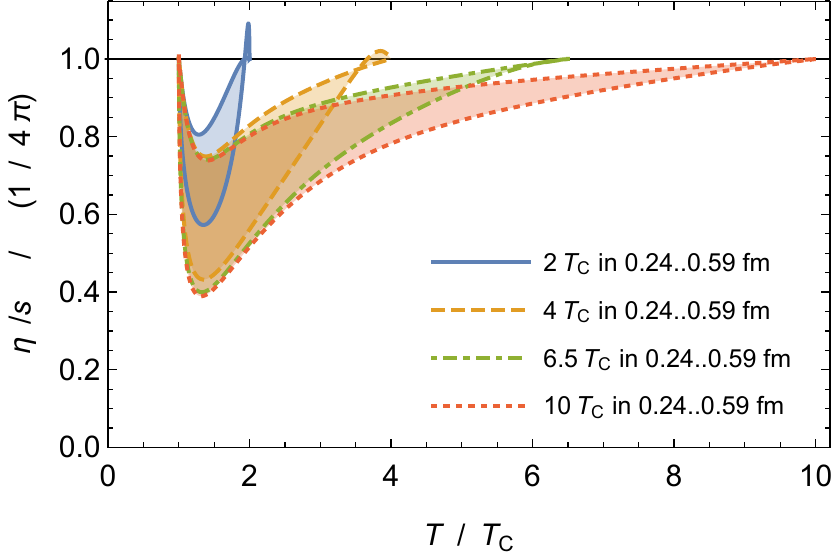}
\caption{\label{fig:etaOfT} 
Dependence of~$\eta/s$ on the instantaneous temperature, $T(t_\text{avg})$. 
The shaded areas indicate the values arising from a sweep over a range of heat-up times for a fixed peak temperature.
}
\end{figure}

In Fig.~\ref{fig:etaOfT}, we display $\eta/s (T)$, which combines the data for $T(t_\text{avg})$ with $\eta/s(t_\text{avg})$.
The curves at heat-up times $\SI{.24}{fm}$ and $\SI{.59}{fm}$ enclose the region which is relevant to heavy-ion collisions. We find a universal behavior: Starting at the critical temperature, the curves bend down, reach their minima around $T=1.3\, T_\text{C}$, and rise toward the equilibrium value.

With increasing peak temperatures, the curves for a given heat-up time become similar, especially near their minima. This implies that the curve at $T=10\, T_\text{C}$ (lower dotted, red) provides a lower bound for $\eta/s$ at each value of $T$. Furthermore, it shows that the influence of the peak temperature on $\eta/s$ saturates. Thus, at large peak temperatures $T\gtrsim 4\, T_\text{C}$, the time span is the dominant out-of-equilibrium parameter which dictates the smallest possible value of $\eta/s$ in a dynamical evolution.

We stress that all the effects discussed here also hold true at non-zero baryochemical potential $\mu$. The analysis of different charge profiles $Q(v)$ in Eq.~\eqref{eq:metric} reproduces the results at vanishing $\mu$. 
A negligible shift of the system response to later $t_\text{avg}$ occurs in line with previous holography results~\cite{Fuini:2015hba,Cartwright:2019opv}. 
These curves are not displayed.

\section{\label{sec:discussion}Discussion}
We have shown, for the first time, that $\eta/s$ receives significant corrections of up to 60\% when evaluated far from equilibrium in a holographic model, cf.~Fig.~\ref{fig:etaOfTime}.   
For all peak temperatures and heat-up times, in the reach of RHIC and LHC, these corrections are of order one, cf.~Fig.~\ref{fig:etaOfT}. 
Holographic models have been shown to capture universal properties of the QGP~\cite{Policastro:2001yc,Kovtun:2004de,Gyulassy:2004zy,Aad:2010bu}.
Hence, we predict universal corrections of order one to $\eta/s$ resulting from far-from-equilibrium physics in strongly coupled plasmas, in particular the QCD quark-gluon plasma. 

\begin{figure}[tb]
\center
\includegraphics[width=.5\linewidth]{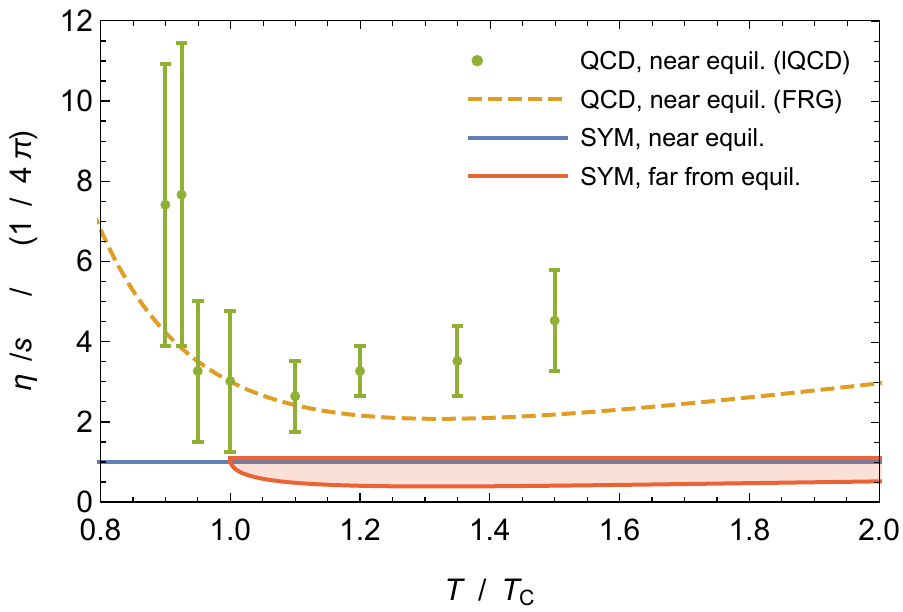}
\caption{\label{fig:etaOfTComparison} 
Temperature-dependence of~$\eta/s$ from our far-from-equilibrium holographic results~(SYM, area enclosed by red solid curve), compared to the holographic near-equilibrium value (SYM, constant blue solid line). 
The theoretical QCD results are computed near equilibrium and stem from the functional renormalization group (FRG, dashed) and lattice QCD (lQCD, circles). 
Data taken from~\cite{Christiansen:2014ypa,Astrakhantsev:2017nrs}.
}
\end{figure}

In Fig.~\ref{fig:etaOfTComparison}, near-equilibrium values of $\eta/s$, as a function of $T$, from different QCD models are compared to values derived from our far-from-equilibrium results. The red shaded area includes all areas in Fig.~\ref{fig:etaOfT}. 
There is a stark contrast: Lattice QCD as well as FRG suggest larger values~\cite{Astrakhantsev:2017nrs,Christiansen:2014ypa}. In agreement with FRG, however, our model predicts a minimum around $1.3\,T_\text{C}$ in all cases, cf.~Fig.~\ref{fig:etaOfT}. 
It is to be expected that our far-from-equilibrium effects are larger than corrections due to a finite number of degrees of freedom and finite 't Hooft coupling~\cite{Kats:2007mq,Brigante:2008gz}. 

Up to now, we have not explained why one could work far from equilibrium in hydrodynamics, which is normally used near equilibrium. 
Hydrodynamics has become the standard tool to interpret experiments at RHIC and LHC~\cite{Kolb:2003dz,Schenke:2011tv,Gale:2013da}. 
Furthermore, it has been shown to work extremely well for theoretical models far from equilibrium~\cite{Chesler:2013lia,Chesler:2008hg,Heller:2013fn,Romatschke:2017vte,Romatschke:2017acs,Strickland:2017kux}. 
The reason for this remains a fundamental question, although theoretical explanations based on resummation and resurgence have been proposed~\cite{Heller:2013fn,Romatschke:2017vte}.
A separate observation is that  Fig.~\ref{fig:etaOfTime} displays an $\eta/s$ which reacts seemingly in a non-linear way to the temperature rise. This non-linear growth may be related to the chaotic evolution of black holes~\cite{Shenker:2013pqa,Maldacena:2015waa}.

Our results show that inclusion of the time-dependence of $\eta/s$ is compulsory in order to extract the properties of QCD matter produced in heavy-ion collisions with high accuracy.
In particular, it will improve the extraction of $\eta/s$ from measured flow coefficients at RHIC and LHC. 
Such a sharpened analysis allows for a deeper understanding of the QGP and opens the possibility for new theoretical insight into time-dependence far from equilibrium at the microscopic level.

\section*{Acknowledgments}
This work was supported, in part, by the U.S.~Department of Energy grant DE-SC-0012447, the Helmholtz International Center for FAIR / LOEWE program (State of Hesse), and the Stiftung Polytechnische Gesellschaft Frankfurt am Main. 
Computational resources were provided by the Frankfurt Institute for Advanced Studies (FIAS). 
We thank 
C.~Cartwright,
T.~Ishii, 
and P.~Nicolini
for discussions. 

%

\providecommand{\href}[2]{#2}\begingroup\raggedright\endgroup

\end{document}